\def\LUM{\:{\rm ergs\:s^{-1}}}

\def\VEL{\:{\rm km\:s^{-1}}}

\def\OIGS{\:{\rm ergs\:cm^{-2}\:s^{-1}\:\AA^{-1}}}
\def\LA{Lyman\thinspace$\alpha$}

\documentclass[12pt,preprint]{aastex}

\begin{document}


\newcommand{\MSOL}{\mbox{$\:M_{\sun}$}}

\newcommand{\EXPN}[2]{\mbox{$#1\times 10^{#2}$}}
\newcommand{\EXPU}[3]{\mbox{\rm $#1 \times 10^{#2} \rm\:#3$}}  
\newcommand{\POW}[2]{\mbox{$\rm10^{#1}\rm\:#2$}}
\newcommand{\SINGLET}[2]{#1$ \lambda $#2}
\newcommand{\DOUBLET}[3]{#1$ \lambda \lambda $#2,#3}
\newcommand{\CHINU}{\mbox{$\chi_{\nu}^2$}}


\title{\bf{WZ Sge: {\it HST} Spectroscopy of the Cooling of the White Dwarf after the 2001
Outburst \footnote{Based on observations made with the NASA/ESA
Hubble Space Telescope, obtained at the Space Telescope Science
Institute, which is operated by the Association of Universities
for Research in Astronomy, Inc., under NASA contract NAS 5-26555.
These observations are associated with proposal \# 9304 and 9459.}
} }

\author{Knox S. Long}

\affil{Space Telescope Science Institute, \\ 3700 San Martin
Drive, \\ Baltimore, MD 21218}

\author{Edward M. Sion}

\affil{Department of Astronomy \& Astrophysics,\\
Villanova University, \\
Villanova, PA, 19085}

\author{Boris T. G\"ansicke}

\affil{Department of Physics \& Astronomy,\\
University of Southampton, \\
Southampton SO17 1BJ UK}

\and

\author{Paula Szkody}

\affil{Department of Astronomy,\\
University of Washington, \\
Seattle, WA 98195}

\begin{abstract}

Following an outburst in 2001 July, the UV flux of WZ Sge has
declined slowly toward pre-outburst levels.  Here, we describe new
1150-1710 \AA\ {\it HST}/STIS spectra of WZ Sge obtained between
2002 April and 2003 March to follow the decline.  A combined
analysis of these and spectra obtained in the fall of 2001 show
that if log g=8.5, then the white dwarf temperature decreased from
23,400 K in 2001 October, shortly after the steady decline of the
system began, to 15,900 K 17 months later. The UV flux in 2003
March was still 2.4 times higher than measured prior to the
outburst and so the system was still recovering from the outburst.
During this period, the shape of the spectrum and the flux from
the system are consistent with a log g=8.5, 0.9 \MSOL\ white dwarf
at the astrometrically-determined distance to WZ Sge. Although the
spectrum from 2001 September resembles that of a white dwarf with
a temperature of 28,200 K, the implied radius is smaller than in
the remainder of the observations. If the entire white dwarf was
visible in 2001 September, then either a second component, such as
the disk emission, was distorting the spectrum, or more likely,
the temperature on the white dwarf photosphere was not uniform
then. The metal lines in the spectra of WZ Sge have weakened with
time. Model fits that allow for material along the line of sight
to the white dwarf photosphere are substantial improvements over
models which assume all the lines arise from the photosphere. This
obviates the need to explain very unusual abundance ratios in the
photosphere and complicates the determination of the rotation rate
of the white dwarf.

\end{abstract}

\keywords{accretion, accretion disks --- binaries: close ---
stars: mass-loss --- novae, cataclysmic variables --- stars:
individual (WZ Sagittae)}

\section{Introduction}

WZ Sge is a cataclysmic variable (CV) with an orbital period of 82
min, close to the period minimum for CVs, which underwent an
outburst for the first time in 22 years in 2001 July.  The
outbursts of WZ Sge and all dwarf novae (DNe) are understood in
the context of a hydrogen-based thermal instability of the
accretion disk which exists around the white dwarf (WD). They are
transitions of the disk from a low temperature, mostly neutral to
a high temperature, mostly ionized state. During outbursts, the
rate of accretion onto the WD in the system is exceptionally high.
WZ Sge is especially interesting because it is the closest example
of a system near the period minimum. Furthermore, the distance is
accurately known from ground-based \citep[43.3$_{-1.5}^{+1.6}$
pc,][]{thorstensen2003} and space-based \citep[43.54$\pm$0.29
pc,][]{harrison2003} parallax measurements. The outburst light
curve in 2001 resembled that of the previous three outbursts that
had been seen. As discussed in detail by \cite{patterson2002}, the
optical light curve consisted of a primary burst lasting 24 days,
followed by a series of minibursts, or echo outbursts, ending in
mid-September, and then a slow steady decline toward a quiescent
level. Some short-period dwarf novae exhibit two types of
outbursts, normal outbursts lasting a few days with amplitudes of
3-5 magnitudes, and less frequent, larger amplitude, longer
duration superoutbursts that are accompanied by characteristic
optical light curve modulations, known as superhumps, with a
period that is slightly different from the orbital period. WZ Sge
does not exhibit two distinct types of outbursts, but the large
amplitude and long duration of the primary outburst, as well as
the identification of surperhumps, suggests that the 2001 and, by
extension, all of the outbursts of WZ Sge should be understood as
superoutbursts.

The 2001 outburst prompted intense study from optical to X-ray
wavelengths. Observations with  {\it FUSE} and {\it HST} during
the primary burst revealed FUV and UV spectra that are fairly
typical of other high inclination DN in outburst, with strong
emission lines of O~VI, N~V and C~IV \citep{knigge2002,long2003}.
This spectrum evolved during and after the primary outburst so
that by the end of the rebrightening phase the FUV and UV spectrum
of WZ Sge appeared to dominated by a slowly cooling WD
\citep{long2003,sion2003}. By November 2001, the WD temperature
was 22,000 K or 19,000 K as determined from the 905-1183 \AA\ {\it
FUSE} spectrum \citep{long2003} or the 1150-1730 \AA\ {\it
HST}/STIS spectrum \citep{sion2003}, respectively. It is not clear
whether the difference in these two temperatures represents a
departure from a simple WD spectrum, a problem in the relative
calibration of {\it FUSE} and STIS, or some combination of these.
In any event, the temperature of the WD was significantly higher
than the temperature of 14,800 K (assuming log g=8) measured with
{\it HST}/GHRS 17 years after the previous outburst
\citep{cheng1997}. In addition to a higher temperature, the
post-outburst spectra show relatively narrow absorption lines
which suggest that the WD is rotating with $v \sin{(i)}$ of
200-300 $\VEL$, if indeed these lines arise in the photosphere of
the WD. \cite{cheng1997} had suggested that the WD was rotating
with a velocity of 1200$_{-300}^{+400} ~\VEL$. A high rotation
rate would be consistent with the fact that WZ Sge often shows
periodic signals with a period of about 28 seconds, which, if due
to the WD spin period, would imply that the WD in WZ Sge is
magnetic and therefore a DQ Her star.\footnote{This explanation
for the periodicity is not universally accepted, in part because
the signal is not always present and in part because it is not at
exactly the same period.  The situation has been further
complicated by the observation of a 15-s periodicity during the
2001 outburst \citep{knigge2002}. The basic alternatives, in
addition to WD rotation, are ZZ Ceti-like WD pulsations or some
kind of oscillation of the inner disk.  \cite{welsh2003} recently
discussed the various possiblities, concluding that ZZ Ceti
pulsations are highly unlikely in WZ Sge due to the fact that the
oscillation period did not change significantly while the WD
cooled by many thousands of degrees.}

Here we describe STIS observations of WZ Sge obtained in 2002 and
2003 to measure the continued cooling of the WD in this system. To
assure uniformity in our analysis of the spectra of WZ Sge, we
have also re-analyzed the spectra obtained in late 2001, beginning
with the observation on 11 September, after the primary outburst
had ended, but during the period of rebrightenings. Our primary
purpose here is to describe the time-averaged spectra from the
observations and to discuss what these spectra reveal about the WD
as a function of time from outburst.

\section{Observations and Qualitative Description of the Data}

All of the observations discussed here were obtained with STIS
using the medium resolution (E140M) echelle grating, the
0.2\arcsec x 0.2\arcsec aperture, and the FUV MAMA operating in a
photon-counting mode. A log of the observations is presented in
Table \ref{table_obs}.  The time between observations in the fall
of 2001 (during our DD observations) was of order a month, whereas
in 2002 and 2003 (during Cycle 11), it has been of order four
months.

With this instrumental setup, the spectra cover the wavelength
range 1150-1710 \AA.
There were no anomalies during the observations themselves. All of
the data were re-calibrated as they were retrieved from the HST
archive in the spring of 2003, and thus our fluxed spectra reflect
the current understanding of instrumental calibration. The spectra
from the individual spectral orders of the echelle grating were
spliced and binned to a resolution of 0.1 \AA\ for the analysis
described here.

As shown in Figure \ref{wzsge_allspec}, WZ Sge faded continuously
throughout the observations, and the effective color temperature
in the 1150-1710 \AA\ band decreased. The rate of change has
slowed with the time from outburst as the system settles back to
its quiescent state. The average flux in the nearly line-free
interval 1425-1525 \AA\ is listed in listed in Table
\ref{table_obs}.  Even though the rate of decline has slowed
considerably, the 1425-1525 \AA\ flux in 2003 March was 2.4 times
the flux recorded with the GHRS of \EXPU{2.5}{-14}{\OIGS} in 1995.

An example of an individual spectrum, the one obtained in 2002
August, about 400 days from outburst maximum, is shown in Figure
\ref{wzsge02aug_lines} with suggested line identifications. This
spectrum is typical of all the spectra far from outburst. The
continuum is dominated by a very broad absorption feature centered
on \LA\ that in a quiescent CV, such as WZ Sge, is generally
interpreted as meaning that the UV emission is dominated by the
WD.  There are a wide range of ionizations represented. The most
extreme example for a single element is carbon; lines of C~I,
C~II, C~III, and C~IV are present. Some lines, such as
\DOUBLET{C~IV}{1548}{1551}, have emission and absorption
components. These emission components, including the emission
component of \LA, are generally assumed to arise from the disk.

The widths of the absorption lines observed in the spectra have
not changed greatly through the course of the observations
reported here.  In order to quantify this statement, we have
measured the profiles of four relatively isolated and largely
unblended features, namely the two components of the Si IV
doublet, \SINGLET{Si~II}{1527} and \SINGLET{Si~II}{1533} using the
interactive fitting routine in SpecView.\footnote{At the
temperatures we will derive later for the WD in WZ Sge, the Si II
lines would be prominent assuming the atmosphere is
metal-enriched. This is not the case for Si~IV; to produce
substantial Si IV in a WD photosphere one requires temperatures of
about 25,000 K. SpecView is a product of the Space Telescope
Science Institute, which is operated by AURA for NASA (see
http://www.stsci.edu/resources/software\_hardware/specview/).} As
shown in Table \ref{table_line}, the line equivalent widths (EW)
of these lines are fairly constant and the measured velocity
widths are of order 300-500 $\VEL$ with no obvious trend in the
data. Only the observation of 11 September 2001 is anomalous, with
larger values for the EW and velocity widths of Si IV than normal.
But this was the observation taken closest to outburst and is the
only observation discussed here where the disk was at least partly
in the outburst state.

\section{Comparison to WD model spectra}

Our primary goal is to determine the time evolution of the spectra
of WZ Sge in the context of WD models.  In order to carry out this
analysis we have constructed a new LTE grid of synthetic spectra
using Ivan Hubeny's TLUSTY200/SYNSPEC46 suite of programs
\citep{hubeny1988,hubeny1995}. The grid has spectra from
11,000-70,000 K for gravities from log g of 7.5 to 9.5. Metal
abundances in the grid are all scaled together from solar values,
and vary from pure H atmospheres to those in which the metals are
10 times the solar value. The spectra have rotation rates $v
\sin{(i)}$ ranging for no rotation to a rotation of 2000 $\VEL$.
Having created the grid, we fit all of the observed spectra in a
uniform fashion, usually, as is described below, using subsets of
the grid.  Specifically, we utilized a standard
$\chi^2$-minimization routine to find the best fits to the data
over the wavelength intervals 1150-1206 \AA, 1226-1233 \AA,
1246-1390\AA,  1410-1538 \AA, and 1560-1710 \AA. Basically,
selecting these ranges was intended to exclude a few prominent
features in the spectra that are not expected in WD photospheres
with temperatures of order 20,000 K or less.  More specifically,
we excluded regions containing the emission features associated
with \LA\ and C~IV that presumably arise from the disk and the
absorption lines of N~V, Si~IV, and C~IV.

As an initial step, we fit all of the spectra to models with log
g=8.5 and the abundances fixed at their solar values. The free
variables for such fits are the WD temperature $T_{wd}$, the
rotation rate of the WD $v\:sin(i)$, and a normalization,
effectively the solid angle of the WD, which we express in terms
of the WD radius $R_{wd}$ at a distance of 43.4 pc.  We chose log
g=8.5 as a fiducial gravity, because recent optical studies of the
disk and secondary indicate that the WD in WZ Sge is fairly
massive ($\ge 0.7 \MSOL$). \cite{skidmore2000} suggest that
$m_{wd} = 0.82\pm0.1$ \MSOL; \cite{steeghs2001} that $m_{wd}
> 0.77 \MSOL $; and \cite{spruit1998} that it is either 0.9 \MSOL\
or 1.2 \MSOL, depending on the details of their analysis. A WD
with log g=8.5 corresponds to a star with a mass of 0.9 \MSOL\ and
a radius of \EXPU{6.5}{8}{cm} \citep{hamada1961}\footnote{Here and
elsewhere we have used the analytic expression for the
Hamada-Salpeter mass-radius (1961) relationship due to Paczynski
as quoted by \cite{anderson1988} to provide $M_{wd}$ and $R_{wd}$
as a function of log g.}, and is in the range suggested by
ground-based observations. The results of the fits are found in
Table \ref{fits}. They reflect our impression from the data that
the effective temperature of WZ Sge is falling in the wavelength
range of the STIS observations, and (due to the fact that the
effective radius of the WD is relatively constant) that the flux
decline that is observed in WZ Sge is simply a reflection of lower
effective temperature.

A more physical approach is to fit the spectra to models for fixed
gravity and variable rotation rate and metallicity. These models
yield relatively good qualitative fits to the spectra, as in shown
in Figures \ref{wzsge02apr_wdg85} and \ref{wzsge03mar_wdg85} for
the 2002 April and 2003 March spectra, and better, but not
statistically acceptable \CHINU. In particular, the overall shapes
of the spectra fit well. The only exception to this rule is at the
shortest wavelengths, where the model fluxes systematically exceed
the data.  The problem at short wavelengths is most likely a
manifestation of a decline in the effective area of STIS at short
wavelengths. Calibration files that correct for this problem exist
currently for the low resolution, but not the echelle grating
modes. It might be possible to correct for this, but in fact the
results of our fits, including the $\CHINU$, do not change
significantly when the 1150-1200 \AA\ region is excluded,  As a
result, we have left the fits as they were. Except for Si~IV, N~V
and C~IV, the lines in the observed spectra also appear in the
models, although not always in the appropriate strengths.

Based on these model fits made with log g=8.5, the WD in WZ Sge
appears to have cooled from 28,200 K in 2001 September to 23,400 K
in 2001 October to 15,900 K in 2003 March. Ignoring the first
observation, when the disk as evidenced by the optical light curve
was still undergoing echo outbursts, the effective radius of the
WD has been fairly constant at \EXPU{6.4}{8}{cm}. This radius is
about that expected for a WD with log g=8.5 (\EXPU{6.5}{8}{cm}).
As expected, the value of $v\:sin(i)$ obtained in the model fits
is fairly similar to that determined from our measurements of
individual line widths; there is not a strong trend in
$v\:sin(i)$, although it is interesting that the measured value is
highest -- 460 $\VEL$--in terms of the fits in 2003 March.

We have also fit the data to grids with log g=8 and log g=9, which
correspond to WDs with masses of 0.6 and 1.2 \MSOL\, respectively.
The radii obtained from the fitting the spectra from 2001 October
through 2003 March is shown in Figure \ref{wdrad}.  The
temperatures derived for the log g=8 models are about 1000 K lower
than for log g=8.5 and the fitted radius, \EXPU{\sim7.5}{8}{cm},
is correspondingly higher. For log g=8, the Hamada-Salpeter radius
is \EXPU{9}{8}{cm}, which is about 30\% larger than the fitted
values, and so the results are not physically consistent, given
the known distance to WZ Sge. For log g=9, $T_{wd}$ is 1000 K
higher than for log g=8.5 and the fitted radius is
less,\EXPU{\sim5.5}{8}{cm}. For log g=9.0, the Hamada-Salpeter
radius is \EXPU{4.2}{8}{cm}, about 25\% smaller than the fitted
value, and strongly suggesting within the context of simple WD
models that the mass must be smaller than 1.2 \MSOL. There are
small differences in \CHINU\ when one compares the various gravity
fits, but no real qualitative difference in the spectra, and thus
we do not believe that one can use the shape of the spectra to
prefer one gravity or another.  If the astrometric distance to WZ
Sge is correct, then log g=8.5 and therefore a mass of about 0.9
\MSOL\ for the WD in WZ Sge is preferred.

Independently of log g, there appears to have been a gradual
decrease in metallicity of the photosphere over time, or at the
very least the strength of the strongest lines relative to that
expected in a solar abundance photosphere. Since metals settle in
the high gravity photosphere on short time scales compared to the
period of our observations, this could reflect either a decrease
in the accretion rate onto the surface of the WD as the system
converts itself to full quiescence, or alternatively that material
along the line of sight to the WD is slowly disappearing (as will
be discussed more fully in Section \ref{abs}).

Although the fits obtained with scaled solar abundances provide
reasonable qualitative fits to the data, there are some
discrepancies, as indicated in Figures \ref{wzsge02apr_wdg85} and
\ref{wzsge03mar_wdg85} for the 2002 April and 2003 March spectra.
Most of the stronger absorption lines in the spectrum of WZ Sge
are due to either C or Si, and in the fits the C lines, especially
C III $\lambda$ 1183 and C II $\lambda$ 1335, are often too weak.

To try to explore the evolution of metal abundance further, we
created a grid of model spectra for log g=8.5 in which the
abundances of C and Si were varied independently from 0.1 to 5
times solar, while the rest of the abundances were maintained at
solar.  As indicated in Table \ref{sic}, somewhat better fits in
terms of \CHINU\ can be obtained in this manner.  Taken at face
value these fits suggest that the Si abundance in the atmosphere
declines by a large fraction, from 3.5 times solar in 2001
September to 0.4 times solar in 2003 March. On the other hand, the
decline in the C abundance is modest, from 5 times solar to $\sim$
3 times solar. The fit to the 2002 April spectrum, shown in Figure
\ref{wzsge02apr_sic}, is typical.  By allowing Si and C to vary
independently the fit to the C II $\lambda$ 1335 line is improved
considerably without affecting the strong feature at 1300 \AA,
which is dominated by Si III.

\section{Absorption along the line of sight to the WD\label{abs}}

Although many of the features in the spectrum of WZ Sge in late
time can be explained in terms of photospheric absorption from a
WD, some of the features have to form along the line of sight to
the photosphere.  Indeed, in choosing the wavelengths ranges to
fit, we explicitly eliminated the N~V and Si~IV regions because
our preliminary fits to the spectrum showed that these lines did
not appear at all at the photospheric temperatures required to fit
the spectrum.  We also eliminated the C~IV region, which shows
both emission and absorption components.  All of this begs the
question of whether the lines we are fitting are actually
photospheric lines. Especially important in this regard is whether
the apparent overabundance of C relative to Si represents the
actual situation in the photosphere of WZ Sge.

Unfortunately, as was discussed by \cite{long2003} in the context
of the {\it FUSE} observations of WZ Sge, the question is not easy
to answer because we do not have a physical model for the material
within the WZ Sge system along the line of sight to the WD.
Nevertheless, as a first step to determine how much overlying
material could affect conclusions regarding the metal abundances
and other characteristics of the WD, we have considered models
consisting of a WD with intervening absorption due to a slab of
material along the line of sight to the WD.  We assume the slab is
in LTE, and can be characterized in terms of a temperature
$T_{slab}$, a total (neutral + ionized) hydrogen column density
$N_H$, a turbulent velocity $v_{slab}$, and a total density $n_H$.
We have calculated the opacity of LTE material using
TLUSTY/SYNSPEC, convolved the opacities with the appropriate
velocity dependent Gaussian to create a smoothed opacity, and
created grids of optical depths as a function of column density.
This approach resembles that used by \cite{horne1994} to model
1150-2500 \AA\ {\it HST} spectra of the high-inclination dwarf
nova OY Car in quiescence in terms of a 16,500 K WD and an ``Fe II
curtain". Our grid included temperatures from 5,000-30,000,
densities from $10^9$ to \POW{13}{cm^{-3}}, turbulent velocities
from 50-500 $\VEL$, and column densities from log($N_H$) of 18 to
22. In fact, because we have assumed LTE, there is a degeneracy
between density and temperature in the models; if a model with a
density of \POW{11}{cm^{-3}} matches the data, then there will be
another model with density \POW{13}{cm^{-3}} that matches equally
well with the same column density but a higher temperature.  Since
we are interested primarily whether departures from solar
abundance ratios are required in the photosphere of WZ Sge, we
have limited the slab models to models in which the material has
solar abundance ratios.

We fit the data from each of the observations to a WD with
intervening absorption.  We used log g=8.5 model spectra for the
WD, allowing the temperature, rotation rate and overall metal
abundance z to vary.  The results for slabs with density
\POW{11}{cm^{-3}} are shown in Table \ref{veil}. In terms of
$T_{wd}$ and $R_{wd}$, the results are very similar to those
derived for pure WD models with the same gravity. This is because
the WD temperature and radius are primarily determined by the \LA\
profile, the overall slope of the continuum and the flux and
because \CHINU\ is also mainly determined by the fit to the
overall continuum.  On the other hand, a comparison of the results
with those obtained from WD models with a variable C and Si
abundance shows that the slab models generally fit the data
better, often considerably better, in a \CHINU\ sense.  An example
of the fitted spectra is shown in Figure \ref{wzsge02apr_veil}.
Thus, our analysis suggests caution in using the STIS data (and by
extension the earlier data) to argue that there are significant
departures from solar abundance ratios in the photosphere of the
WD in WZ Sge.

\section{The Quiescent GHRS spectrum}

As noted earlier, WZ Sge had been observed in 1995 with the GHRS
\citep{cheng1997}. A comparison of the 2003 March STIS spectrum to
the GHRS spectrum, which was obtained with the G140L grating at a
resolution of $\sim$2000, is shown in Figure \ref{wzcomp}. To
emphasize the differences in shape, the GHRS spectrum has been
rescaled to the 2003 March flux in the wavelength range 1450-1500
\AA. The GHRS spectra cover a smaller spectral range -- 1237-1523
\AA\ -- than the post-outburst STIS spectra. Inspection of this
figure shows a much broader Lyman $\alpha$ profile in 1995
September than in 2003 March.  This is an indication that in 2003,
the WD had not reached the quiescent temperature of the WD. On the
other hand, nearly all of the features in the 2003 March spectrum
are also seen in the 1995 September spectrum, and the shapes,
though not necessarily the depths, are rather similar.

In order to make the comparison with the STIS spectra more
quantitative, we have refit the GHRS spectrum with our spectral
models. To provide the most direct comparison possible with the
STIS data, we fit the regions 1237-1233 \AA, 1246-1390 \AA, and
1410-1523 \AA. The best log g=8.5 variable metallicity fit yields
a WD with a temperature of 14,400 K, a WD radius of
\EXPU{5.6}{-8}{cm}, and a \CHINU\ of 2.48. This is approximately
1500 K cooler than the value of 15,900 K obtained from similar
fits to the data from 2003 March. Thus there is good evidence that
WZ Sge is not yet at its quiescent temperature, and will continue
to cool. The best fit value of the metallicity is quite large,
8.1, and the value of $v\:sin(i)$ is  the highest value in our
grid, 2000 $\VEL$. The metallicity and the value of $v\:sin(i)$
are correlated; as the metallicity increases, $v\:sin(i)$ also
must increase to match the depth of lines. The radius implied by
the best fit to the GHRS data is about 13\% smaller than found in
the STIS data. But this result is quite sensitive to the actual
temperature. If we constrain the WD radius to be \EXPU{6.5}{8}{cm}
and fit the GHRS spectrum, then the implied temperature is 14,200
K, only 200 K lower than when the radius is
unconstrained.\footnote{While in WZ Sge there is no evidence that
the emitting area of the WD has changed significantly with time
from outburst, the situation is different in some other systems,
notably AL Com.  There, \cite{szkody2003} find that the UV
spectral shape is constant but the flux varies by a factor of two
in observations separated by 4.5 years in one interoutburst
interval.  Since both spectra can be fit to a WD with a
temperature about 16,000-17,000 K, this suggests that the emitting
area has changed, or that there is a second and variable component
in the UV emission in that system.}

\cite{cheng1997} had used log g=8 models and found a temperature
of 14,800 K; for log g=8, our models in which all of the
abundances scale together suggest a somewhat lower temperature
13,400 K. This result does not depend on whether we use the
intervals selected to match as closely as possible those used in
the STIS data, or the intervals used by \cite{cheng1997}.  We
suspect that the discrepancy in temperatures is due to slight
differences in the model inputs, which leads to a slight
difference in the temperature scale, or possibly updates in the
calibration files or calibration pipeline used when we reprocessed
the GHRS data.
While it is not entirely clear whether our new temperature is
better in an absolute sense, we do think the optimum approach to
minimizing the model inputs is to use a single model grid to
intercompare the STIS and GHRS data.

\section{The 2001 September Spectrum Revisited}

Regardless of whether the bulk of the metal lines are created in a
WD photosphere or by absorbing material, all of model fits we have
discussed thus far indicate that the radius of the WD was lower in
2001 September than later.  That the 2001 September 11 spectrum is
different is not especially surprising. Although this observation
occurred after the main outburst which ended on 16 August, it
preceded the last miniburst in the optical light curve on 2001
September 13.\footnote{The relative timing of this observation and
the mini-outbursts are shown in Figure 1 of \cite{long2003}} The
nature of the mini-outbursts are not well understood.  They are
common in WZ Sge systems, but no system that exhibits mini-bursts
has been observed in enough detail at UV wavelengths to know how
UV spectra are affected.  The optical rebrightenings are assumed
to be some kind of disk outburst. \cite{osaki2001} have shown that
one way to produce this type of behavior is by variations in
viscosity that might be connected to a decay in the magnetic field
of the disk.  Alternatively, \cite{hameury2000} have suggested
that irradiation of the secondary during the primary outburst
increases the mass transfer rate from the secondary over the
normal quiescent state and results in mini-outbursts. The optical
hot spot is eclipsed in WZ Sge; \cite{patterson2002} has used
brightness of the hot spot to suggest that the mass transfer rate
from the secondary $\dot{m}_{2}$ is in the range
\EXPU{0.7-2.0}{16}{gm~s^{-1}}, or \EXPU{1.1-3.2}{-10}{\MSOL~yr^{
-1}}.  This is large enough to produce weak DN outbursts.

The basic anomaly in our simple analysis of the 2001 September is
that the radius of the WD appears smaller than in later
observations. One might imagine that the disk penetrates to the WD
surface and therefore obscures a portion of the surface.  However,
an alternative is that a portion of the surface is actually cooler
than indicated by all of the model fits in which the photospheric
temperature is constant over the entire surface.  Analyses of the
spectra of various other cataclysmic variables, including U Gem
\citep{long2003} and VW Hyi \citep{gaensicke1996}, suggest that
the temperature on the surface of the WD is not always uniform.
Possible explanations for this have involved on-going accretion
that heats a portion of the WD surface and/or emission from an
accretion belt, a ring of rapidly rotating matter in the surface
layers of the WD created from accretion of high angular momentum
disk material \citep{kippenhahn1978, long1993}. To see whether
this model is a likely explanation in the case of WZ Sge, we have
considered two-temperature fits to the spectrum. Specifically, we
tried to fit the data in terms of two log g=8.5 components, each
with separate temperatures, metal abundances scaled to solar, and
rotation rate. The best fit in this case has a 29,000 K component,
with 5 times solar abundances rotating at 390 $\VEL$ and an 18,500
K component with 9.9 times solar abundances rotating at 500
$\VEL$.  The hotter component occupies 57\% of the surface and the
total radius in this case is \EXPU{6.5}{8}{cm}. Thus, in
principle, one can reconcile the radius problem with a
multi-temperature photosphere. On the other hand, \CHINU\ is 12.2,
only slightly improved over the identical model with one
component.

It is very unlikely that the disk totally dominates the spectrum
in 2001 September.  Firstly, \cite{sion2003} showed that the best
fit disk models, in which $m_{wd}$, $\dot{m}_{disk}$, and
inclination were varied, lead to unrealistic parameters for the
accretion. Secondly, the flux observed in September was only 1.5
times that in October.  It is very difficult to believe that the
WD was intrinsically fainter in September. So unless the WD was
somehow obscured in September, it should comprise 2/3 of the
observed flux. But we also note that if the disk was close to
steady state and the disk accretion rate was \POW{16}{g~s^{-1}},
or similar to that suggested by the \cite{patterson2002} estimate
of $\dot{m}_{2}$, then the expected flux at 1500\AA\ from the disk
would be \EXPU{\sim5}{-13}{\OIGS}. This is more than is actually
observed. Nevertheless, it is another indication that there could
be a significant disk contribution to the 2001 September spectrum.

So we have attempted to test for this. For the disk, we used the
models based on summing appropriately-weighted stellar and
rotationally-broadened stellar spectra previously constructed by
\cite{long2003} for their analysis of the {\it FUSE} outburst
spectra. The disk variables that were fit were the disk accretion
rate and a normalization.  The models were constructed assuming
the known distance to WZ Sge and an inclination of 75$\degr$. The
normalization is expressed as the ratio of the contribution from
the disk component in the fit to that expected given the distance.
To be a self-consistent representation of the disk contribution,
the normalization should have a value close to 1. To model the WD,
we used our standard log g=8.5 grid in which the variables are the
WD temperature, the overall metal abundance, the rotation rate,
and the WD radius. The best fit was for a model with a negligible
disk contribution (and low normalization compared to 1) and a WD
with parameters of our earlier model fit for a pure WD. We also
attempted fits in which we constrained the normalization for the
disk contribution to be 1 and set the radius of the WD to be
\EXPU{6.4}{8}{cm}. In this case, the best fit occurs with
$\dot{m}_{disk}$ at the minimum value in the grid, or
\POW{15}{gm~s^{-1}}.  Both fits suggests that the contribution of
the disk to the 2001 September spectrum was small. It appears more
likely that the problem of a small WD radius in simple WD model
fits to the 2001 spectrum is due to a non-uniform temperature
distribution than disk emission. The main difficulty in drawing a
firm conclusion is our lack of confidence that the disk model
spectra reflect the shape of the actual disk contribution in WZ
Sge during the time period of the echo outbursts.

\section{Discussion}

Prior to the outburst, the WD in WZ Sge had a luminosity of about
\POW{31}{ergs~s^{-1}}, based on our determination of the
temperature. The effect of the outburst on the WD was significant.
The luminosity of the WD was at least 7 times greater after the
outburst (13 times greater if the disk contribution to the 2001
September spectrum was small). The time evolution of the
luminosity and temperature of the WD is shown graphically in
Figure \ref{coolfit}.  Assuming a $m_{wd}$ of 0.8 \MSOL,
\cite{long2003}{ estimated that the disk accretion rate in WZ Sge
near the peak of the outburst was \EXPU{8.5}{-10}{\MSOL~yr^{-1}}
implying a disk luminosity of \EXPU{4}{33}{erg~s^{-1}}.  The
primary outburst lasted 24 days and therefore a crude estimate of
the energy radiated by the disk during the outburst is
\EXPU{8}{39}{ergs}.  A simple estimate of the excess energy
radiated by the WD after the outburst can be obtained by fitting
the luminosity excess from 2001 October through 2003 March to an
exponential; the time constant for such a fit is 177 days and the
excess energy is \EXPU{1.2}{39}{ergs}. In fact this may be an
underestimate since the luminosity is not falling as fast as an
exponential toward the end of the observations.  Nevertheless a
value of \EXPU{1.2}{39}{ergs} corresponds to 15\% of the energy
lost from the disk, or 7.5\% of the total accretion energy if half
is radiated by the boundary layer. A variety of explanations have
been invoked to explain the fact that the WD is heated by the
effects of the outburst.  These have included direct heating of
the photosphere \citep{pringle1988}, a somewhat elevated accretion
rate after the outburst \citep{long1993}, the development and
subsequent decay of a rapidly rotating region of the WD
\citep{kippenhahn1978,long1993}, and finally the slow relaxation
of the internal structure of the WD due to the fact that, in the
case of WZ Sge, \POW{23}{gm} of material has been added to the WD
\citep{sion1995a}.

In a similar analysis of normal and superoutbursts of VW Hyi,
\cite{gaensicke1996} found  time constants of 2.8 and 9.8 days and
excess energies radiated by the WD of 0.2 and
\EXPU{1.2}{38}{ergs}, respectively. They estimated about 1\% of
the outburst energy ($\lambda >$ 912 \AA) goes into heating of the
WD in both cases, considerably less than our comparable estimate
of 15\% for WZ Sge. This may indicate that the longer, lower
luminosity outburst in WZ Sge is more efficient in transferring
energy and/or mass to the WD. Alternatively, since the timescales
associated with various mechanisms that produce excess luminosity
in the WD differ, the full impact of the outburst on the WD may
only be apparent in a system, like WZ Sge, with a long
interoutburst interval.  Indeed there is a suggestion that several
of the mechanisms are operating, since, as indicated by the
departures from a simple exponential cooling law shown in Figure
\ref{coolfit}; the luminosity decay timescale is more rapid in
2001 than in late 2002 and 2003.   A detailed attempt to model the
cooling of the WD and to determine the importance of the
mechanisms contributing to the heating and cooling of the WD will
be presented by \cite{godon2003}.

One of our primary goals in following the evolution of the
spectrum of WZ Sge has been to obtain an accurate measurement of
the rotation rate of the system.  For a rotation period of 29 s,
an inclination of 75\degr and a radius of \EXPU{6}{8}{cm}, the
expected $v \sin{(i)}$ is 3200 $\VEL$, even larger than the value
of 1200$_{-300}^{+400} ~\VEL$ measured by
\cite{cheng1997}.\footnote{The velocity could be reduced to 1600
$\VEL$ if the WD is a two-pole system} Most of our model fits show
some evolution of rotation rate with time, with the highest
rotation rates measured in 2003 March. For example, the fits in
which Si and C are allowed to vary independently show, as
indicated in Table \ref{sic}, $v \sin{(i)}$ of 230 and 310 $\VEL$
in 2001 October and November rising to 570 and 680 $\VEL$ in 2002
November and 2003 March. Even the models
 with an absorbing slab (Table \ref{veil}) show $v \sin{(i)}$ of about 250
$\VEL$ in 2001 October and November but 850 $\VEL$ in 2002
November and 2003 March.  This could indicate that a rapidly
rotating WD is gradually being revealed. However, at present the
evidence for rapid rotation, by which we mean evidence rotation
rates comparable to that indicated by the 29 s periodicity, is
weak. There are two main problems. First, while our model fits
reproduce the overall continuum well qualitatively, they are still
poor in a statistical (\CHINU) sense and hence error estimation is
difficult. Second, there is no clean way to separate the portions
of the line profiles associated with the WD and those associated
with intervening material and this is necessary to actually
measure  $v \sin{(i)}$. Our best bet is to wait and hope that the
line of sight absorption will decrease, and for this reason we
have delayed the last observation of WZ Sge in this program to
2004. However, we know from the GHRS observations that some line
of sight absorption will remain, and that if the rotation rate is
as high as 3200 $\VEL$, it will be hard to measure accurately. The
crux of the problem is that one would like to measure shapes of
individual lines to eliminate the correlations and model
dependencies associated with global fits to the data, but that
this is difficult since the lines will be shallow features in a
wavelength-dependent continuum.

The last outburst of WZ Sge occurred in 1978. The system was
observed both in outburst and in decline, although unfortunately
the first ultraviolet observation in the decline phase occurred
about 180 days after the outburst began. The character of the
spectrum was similar to that discussed here. \cite{slevinsky1999}
found that the temperature of the WD in WZ Sge, assuming log g=8,
declined from 20,500 K with an e-folding time of 690 days to
15,400 K. Our analysis would suggest a temperature 1000-2000 K
lower at 180 days if log g=8, but this difference is as likely due
to the difference in instrumental calibrations and models as to
the differences in the outbursts.  They also found in the context
of simple WD photosphere models that the metal abundance declined
with time, and that there were very elevated C abundances and very
anomalous abundance ratios. When we perform a similar analysis of
the STIS data, we obtain similar results. However, as we have
pointed out, absorption from material along the line of sight to
the WD in the 75\degr\ inclination system that comprises WZ Sge
appears to compromise the evidence for anomalous abundance ratios,
including elevated C compared to other metals.

\section{Summary}

The outburst of WZ Sge in 2001 has and will continue to provide a
unique opportunity for advancing our understanding of how a WD
responds to a huge change in accretion rate. It is unique because
the long (almost 23 year) interoutburst interval makes it easy to
isolate effects of the outburst on the WD, and because WZ Sge's
proximity (43.4 pc) makes high quality UV spectra straightforward
to obtain. As a result, we are close to unravelling the system,
and in any event the following seems clear:

\begin{itemize}
\item The WD in WZ Sge, which had a temperature of 14,400 K (using
our analysis) prior to the outburst, was heated to at least 23,300
K and possibly as high as 28,000 K by the outburst (assuming log
g=8.5). \item Seventeen months after the outburst, the WD had
cooled to 15,900 K, but was still feeling the effects of the
outburst. \item Our WD model fits are self-consistent if the WD
has a mass of about $\sim0.9~\MSOL$, that is if the radius is
\EXPU{6.4}{8}{cm} and log g=8.5 and if the distance is the
astrometric value of 43.4 pc. This mass estimate confirms mass
estimates obtained by \cite{steeghs2001} and by
\cite{skidmore2000} by completely different methods. It is
unlikely that the mass is as large as 1.2\MSOL, the higher of the
two masses suggested by \cite{spruit1998}, or as low as 0.6\MSOL.
\item The strength of the metal lines has declined with time,
which is consistent either with a decrease in the metallicity of
the photosphere or a decrease in the amount of material along the
line of sight to the WD. An analysis of the spectra assuming the
bulk of the lines originate in the photosphere would indicate that
the WD is rotating at 300-600 $\VEL$ and that there is strong
overabundance of C relative to Si in the photosphere. \item
However, at present, it is not possible to prove that the bulk of
the lines are from the photosphere. As a result, there is no
requirement to explain a very anomalous set of abundance ratios in
WZ Sge. It is also not possible to conclude whether the WD in WZ
Sge is rotating at the rate needed to explain the 29 s
periodicity, something that is really required to determine that
WZ Sge is a DQ Her system.
\end{itemize}

As noted, we have concentrated here on obtaining the basic
parameters of the WD in WZ Sge as a function of time following the
2001 outburst, and only provide a brief sketch of some of the
implications of the time history.  A complete discussion of the
cooling of the WD, including comparisons between the observed
cooling curve and evolutionary accreting WD model simulations with
time-variable accretion, taking into account the WD rotation,
boundary layer irradiation and compressional heating will be
reported by \cite{godon2003}

\acknowledgments{This work was supported by NASA through grant
G0-09459 from the Space Telescope Science Institute, which is
operated by AURA, Inc., under NASA contract NAS5-26555.  BTG was
supported by a PPARC Advanced Fellowship.}


\pagebreak

\figcaption[fig_allspec]{The {\it HST}/STIS spectra of WZ Sge
obtained as the source faded between 2001 September and 2003
March.  The various spectra are all plotted on the same scale and
have been boxcar smoothed to a resolution of about 1 \AA\ in order
to improve the visual appearance of the figure. Spectra are
plotted alternatively in black and grey. \label{wzsge_allspec}}

\figcaption[fig_wzsge02aug_lines]{The spectrum of WZ Sge obtained
on 2002 August 27, almost 400 days after optical maximum with
suggested identifications for the more prominent line features.
\label{wzsge02aug_lines}}

\figcaption[wzsge02apr_wdg85]{The 2002 April spectrum compared to
the best-fitting log g=8.5 model when the overall metallicity and
rotation rate has been allowed to vary. In the upper panel, the
observed and model spectra are plotted in black and red,
respectively. Data that were excluded from the fitting are plotted
in green. In the lower panel, the difference between the observed
spectrum and the fitted spectrum is shown in black; the two blue
lines in this panel indicate the statistical error.
\label{wzsge02apr_wdg85}}

\figcaption[wzsge03mar_wdg85]{The 2003 March spectrum  compared to
the best-fitting log g=8.5 model when the overall metallicity and
rotation rate has been allowed to vary. \label{wzsge03mar_wdg85}}

\figcaption[wdrad]{A comparison between the radii derived for log
g of 8.0, 8.5, and 9.0 variable z model fits to the spectra
obtained between 2001 October and 2003 March assuming the
astrometric distance to WZ Sge and that expected from a
Hamada-Salpeter relation. The models with log g of 8.0, 8.5 and
9.0 are plotted as open circles, filled circles, and open
rectangles, respectively. There is consistency if log g=8.5, or
the mass of the WD is about 0.93 \MSOL. \label{wdrad}}

\figcaption[wzsge02apr_sic]{The spectrum obtained in 2002 April
compared to the best-fitting log g=8.5 model, when the
temperature, rotation, and  Si and C abundances have been allowed
to vary independently.  Other elements were fixed at solar
abundances in these models.\label{wzsge02apr_sic}}

\figcaption[wzsge02apr_veil]{The spectrum obtained in 2002 April
compared to a model intended to mimic a WD with an absorbing slab
along the line of sight. Here, in the upper panel the unabsorbed
WD model spectrum is shown in blue, while the full model is shown
in red.  The main effect of the absorbing slab is to provide a
better match the absorption features in the spectrum.
\label{wzsge02apr_veil}}


\figcaption[wzcomp]{A comparison between the spectrum (in grey)
obtained with the GHRS in 1995 prior to the outburst with the
spectrum (in black) in 2003 March, 607 days after the 2001
outburst. The GHRS spectrum has been rescaled upward by a factor
of 2.4 to match the flux in the 1450-1500 \AA\ region of the STIS
spectrum. The STIS spectrum has been box-car smooth to approximate
the resolution of the GHRS spectrum. The broad feature near 1400
\AA\ in the GHRS spectrum is due to quasimolecular hydrogen.
\label{wzcomp}}


\figcaption[coofit]{Time evolution of the luminosity (upper panel)
and temperature (lower panel) of the WD in WZ Sge after the 2001
outburst. The solid lines show the measured values; the dotted
lines show fits to the measured values using the data from 2001
October onward; and the dashed lines are values derived from the
pre-outburst GHRS observations. For the luminosity, the functional
form of the fitted curve is \EXPN{1.05}{32} exp(-$\tau$/177 days)
+ \EXPU{1.0}{31}{\LUM}; for the temperature, it is 12,200
exp(-$\tau$/242 days) + 14,300 K. Note that there is a substantial
excess over an exponential fit for the 2001 September observations
that may indicate the effects of ongoing accretion during the time
period when echo outbursts were occurring. \label{coolfit}}

\clearpage

\begin{center}
\begin{deluxetable}{cccc}
\tablecaption{Observation Log }
\tablehead{\colhead{Date} &
 \colhead{Time~Past~Max.$^{a}$} &
 \colhead{Exposure} &
 \colhead{1425-1525~\AA~Flux}
\\
\colhead{~} &
 \colhead{(Days)} &
 \colhead{(s)} &
 \colhead{($\POW{-14}{\OIGS}$)}
}
\scriptsize
\tablewidth{0pt}\startdata
2001-09-11 &  ~49.1 &  2330 &  38.6 \\
2001-10-10 &  ~78.6 &  2330 &  26.5 \\
2001-11-10 &  109.5 &  2330 &  21.1 \\
2001-12-11 &  140.0 &  2330 &  17.2 \\
2002-04-15 &  265.4 &  5150 &  10.3 \\
2002-06-05 &  316.3 &  5150 &  ~9.1 \\
2002-08-27 &  399.7 &  5150 &  ~8.0 \\
2002-11-01 &  465.1 &  5150 &  ~6.6 \\
2003-03-23 &  607.2 &  5150 &  ~6.0 \\
\tablenotetext{a}{ Outburst maximum assumed to occur 2001 July 24
0 hr UT or JD 2452114.5.}
\enddata
\label{table_obs}
\end{deluxetable}
\end{center}

\begin{center}
\begin{deluxetable}{lcccccccc}
\tablecolumns{9} \tablewidth{0pt}

\tablecaption{Line Strengths and Widths}

\tablehead{

\colhead{Date} & \multicolumn{4}{c}{Si~IV} &
\multicolumn{4}{c}{Si~II}
\\

\multicolumn{1}{c}{~} & \multicolumn{2}{c}{$\lambda$1393.8} &
\multicolumn{2}{c}{$\lambda$1402.8} &
\multicolumn{2}{c}{$\lambda$1526.7} &
\multicolumn{2}{c}{$\lambda$1533.4}
\\

\colhead{~} & \colhead{EW} & \colhead{FWHM} & \colhead{EW} &
\colhead{FWHM} & \colhead{EW} & \colhead{FWHM} & \colhead{EW} &
\colhead{FWHM}

\\

\colhead{~} & \colhead{(\AA)} & \colhead{($\VEL$)} &
\colhead{(\AA)} & \colhead{($\VEL$)} & \colhead{(\AA)} &
\colhead{($\VEL$)} &  \colhead{(\AA)} & \colhead{($\VEL$)}

}

\scriptsize \tablewidth{0pt} \startdata
2001-09-11 &  4.6 &  1251 &  4.0 &  1142 &  0.8 &  340 &  0.7 &  310 \\
2001-10-10 &  2.2 &  500 &  1.9 &  432 &  1.4 &  361 &  1.6 &  380 \\
2001-11-10 &  2.1 &  465 &  1.8 &  402 &  1.4 &  371 &  1.6 &  380 \\
2001-12-11 &  2.0 &  459 &  1.7 &  381 &  1.4 &  361 &  1.6 &  390 \\
2002-04-15 &  1.8 &  390 &  1.4 &  316 &  1.0 &  329 &  1.2 &  326 \\
2002-06-05 &  1.8 &  416 &  1.6 &  372 &  1.4 &  371 &  1.6 &  380 \\
2002-08-27 &  1.7 &  454 &  1.3 &  368 &  0.9 &  345 &  1.2 &  453 \\
2002-11-01 &  1.4 &  433 &  1.0 &  320 &  0.7 &  355 &  1.0 &  461 \\
2003-03-23 &  1.6 &  529 &  1.1 &  416 &  0.7 &  361 &  0.8 &  385 \\
\enddata
\label{table_line}
\end{deluxetable}
\end{center}

\begin{center}
\begin{deluxetable}{ccccccc}
\tablecaption{WD Model Fits }
\tablehead{\colhead{Model~type} &
 \colhead{Observation} &
 \colhead{$\CHINU$} &
 \colhead{$R_{wd}$\tablenotemark{a}} &
 \colhead{T} &
 \colhead{v~sin(i)} &
 \colhead{Metallicity}
\\
\colhead{~} &
 \colhead{~} &
 \colhead{~} &
 \colhead{(\POW{8}{cm})} &
 \colhead{(1000$\degr$~K)} &
 \colhead{($\VEL$)} &
 \colhead{~}}
\scriptsize
\tablewidth{0pt}
\startdata
log(g)~=~8.5,~Solar &  2001-Sept. &  16.4 &  5.9 &  26.2 &  310 &  - \\
~ &  2001-Oct. &  ~9.9 &  6.5 &  22.6 &  180 &  - \\
~ &  2001-Nov. &  ~7.5 &  6.5 &  21.4 &  260 &  - \\
~ &  2001-Dec. &  ~4.9 &  6.4 &  20.5 &  180 &  - \\
~ &  2002-Apr. &  ~5.7 &  6.5 &  17.9 &  360 &  - \\
~ &  2002-Jun. &  ~4.8 &  6.5 &  17.4 &  310 &  - \\
~ &  2002-Aug. &  ~3.7 &  6.6 &  16.8 &  430 &  - \\
~ &  2002-Nov. &  ~3.0 &  6.3 &  16.5 &  470 &  - \\
~ &  2003-Mar. &  ~2.5 &  6.3 &  16.1 &  560 &  - \\
log(g)~=~8.0,~Variable~z &  2001-Sept. &  12.8 &  6.0 &  27.1 &  450 &  5.4 \\
~ &  2001-Oct. &  ~9.7 &  7.1 &  22.5 &  240 &  2.4 \\
~ &  2001-Nov. &  ~7.7 &  7.3 &  20.7 &  300 &  1.9 \\
~ &  2001-Dec. &  ~5.2 &  7.2 &  19.5 &  230 &  1.5 \\
~ &  2002-Apr. &  ~6.0 &  7.8 &  16.7 &  360 &  0.9 \\
~ &  2002-Jun. &  ~5.0 &  7.7 &  16.4 &  340 &  1.2 \\
~ &  2002-Aug. &  ~3.8 &  7.9 &  15.6 &  380 &  0.7 \\
~ &  2002-Nov. &  ~3.0 &  7.6 &  15.3 &  420 &  0.6 \\
~ &  2003-Mar. &  ~2.5 &  7.7 &  14.9 &  460 &  0.4 \\
log(g)~=~8.5,~Variable~z &  2001-Sept. &  12.7 &  5.5 &  28.2 &  430 &  5.0 \\
~ &  2001-Oct. &  ~9.2 &  6.2 &  23.4 &  230 &  2.1 \\
~ &  2001-Nov. &  ~7.1 &  6.2 &  22.1 &  290 &  1.9 \\
~ &  2001-Dec. &  ~4.8 &  6.3 &  20.7 &  190 &  1.4 \\
~ &  2002-Apr. &  ~5.7 &  6.4 &  18.1 &  360 &  1.1 \\
~ &  2002-Jun. &  ~4.8 &  6.5 &  17.4 &  320 &  1.2 \\
~ &  2002-Aug. &  ~3.6 &  6.7 &  16.7 &  380 &  0.8 \\
~ &  2002-Nov. &  ~2.8 &  6.4 &  16.3 &  420 &  0.6 \\
~ &  2003-Mar. &  ~2.4 &  6.4 &  15.9 &  460 &  0.5 \\
log(g)~=~9.0,~Variable~z &  2001-Sept. &  12.4 &  5.0 &  29.3 &  390 &  4.8 \\
~ &  2001-Oct. &  ~8.7 &  5.5 &  24.7 &  210 &  2.2 \\
~ &  2001-Nov. &  ~6.7 &  5.4 &  23.4 &  280 &  1.9 \\
~ &  2001-Dec. &  ~4.5 &  5.3 &  22.2 &  180 &  1.5 \\
~ &  2002-Apr. &  ~5.4 &  5.5 &  19.3 &  340 &  1.1 \\
~ &  2002-Jun. &  ~4.8 &  5.5 &  18.6 &  290 &  1.2 \\
~ &  2002-Aug. &  ~3.7 &  5.6 &  17.9 &  370 &  0.8 \\
~ &  2002-Nov. &  ~2.9 &  5.4 &  17.4 &  380 &  0.6 \\
~ &  2003-Mar. &  ~2.5 &  5.4 &  17.1 &  460 &  0.5 \\
\enddata
\tablenotetext{a}{Here and in other tables, for a distance of 43.3
pc} \label{fits}
\end{deluxetable}
\end{center}

\begin{center}
\begin{deluxetable}{ccccccc}
\tablecaption{WD Model Fits with variable Si and C abundance }
\tablehead{\colhead{Observation} &
 \colhead{$\CHINU$} &
 \colhead{$R_{wd}$} &
 \colhead{T} &
 \colhead{v~sin(i)} &
 \colhead{Si} &
 \colhead{C}
\\
\colhead{~} &
 \colhead{~} &
 \colhead{(\POW{8}{cm})} &
 \colhead{(1000$\degr$~K)} &
 \colhead{($\VEL$)} &
 \colhead{~} &
 \colhead{}
}
\scriptsize
\tablewidth{0pt}\startdata
2001-Sept. &  12.5 &  5.1 &  28.0 &  410 &  3.5 &  5.0 \\
2001-Oct. &  ~9.0 &  6.2 &  23.3 &  230 &  1.9 &  3.6 \\
2001-Nov. &  ~6.6 &  6.2 &  22.0 &  310 &  1.5 &  4.6 \\
2001-Dec. &  ~4.5 &  6.3 &  20.6 &  240 &  0.9 &  3.0 \\
2002-Apr. &  ~4.9 &  6.4 &  18.1 &  440 &  0.6 &  3.5 \\
2002-Jun. &  ~4.3 &  6.5 &  17.5 &  360 &  0.7 &  2.9 \\
2002-Aug. &  ~3.0 &  6.6 &  16.9 &  530 &  0.5 &  3.3 \\
2002-Nov. &  ~2.4 &  6.3 &  16.5 &  570 &  0.4 &  3.1 \\
2003-Mar. &  ~2.0 &  6.3 &  16.2 &  680 &  0.4 &  3.5 \\
\enddata
\label{sic}
\end{deluxetable}
\end{center}

\begin{center}
\begin{deluxetable}{ccccccccc}
\tablecaption{Model with a WD and Absorbing Slab }
\tablehead{\colhead{Observation} &
 \colhead{$\CHINU$} &
 \colhead{$R_{wd}$} &
 \colhead{$T_{wd}$} &
 \colhead{v~sin(i)} &
 \colhead{z} &
 \colhead{$T_{slab}$} &
 \colhead{$N_H$} &
 \colhead{$v_{turb}$}
\\
\colhead{~} &
 \colhead{~} &
 \colhead{(\POW{8}{cm})} &
 \colhead{(1000$\degr$~K)} &
 \colhead{($\VEL$)} &
 \colhead{~} &
 \colhead{(1000$\degr$~K)} &
 \colhead{(cm$^{-2}$} &
 \colhead{($\VEL$)}
}
\scriptsize
\tablewidth{0pt}\startdata
2001-Sept. &  7.4 &  5.1 &  29.1 &  1200 &  2.4 &  11.5 &  21.3 &  240 \\
2001-Oct. &  5.0 &  6.2 &  23.2 &  ~240 &  0.8 &  ~7.2 &  19.5 &  220 \\
2001-Nov. &  4.1 &  6.6 &  21.1 &  ~260 &  0.4 &  ~6.7 &  19.1 &  430 \\
2001-Dec. &  2.9 &  6.5 &  20.1 &  ~140 &  0.3 &  ~6.4 &  19.0 &  450 \\
2002-Apr. &  3.7 &  6.7 &  17.6 &  ~510 &  0.3 &  ~6.5 &  19.0 &  420 \\
2002-Jun. &  2.9 &  6.2 &  17.2 &  ~730 &  0.4 &  ~6.7 &  19.1 &  150 \\
2002-Aug. &  2.4 &  6.7 &  16.6 &  ~830 &  0.4 &  ~6.6 &  18.9 &  260 \\
2002-Nov. &  2.1 &  6.3 &  16.3 &  ~850 &  0.3 &  ~6.5 &  18.8 &  260 \\
2003-Mar. &  2.0 &  6.4 &  15.9 &  ~850 &  0.3 &  ~6.5 &  18.7 &  260 \\
\enddata
\label{veil}
\end{deluxetable}
\end{center}

\pagestyle{empty}
\begin{figure}
\plotone{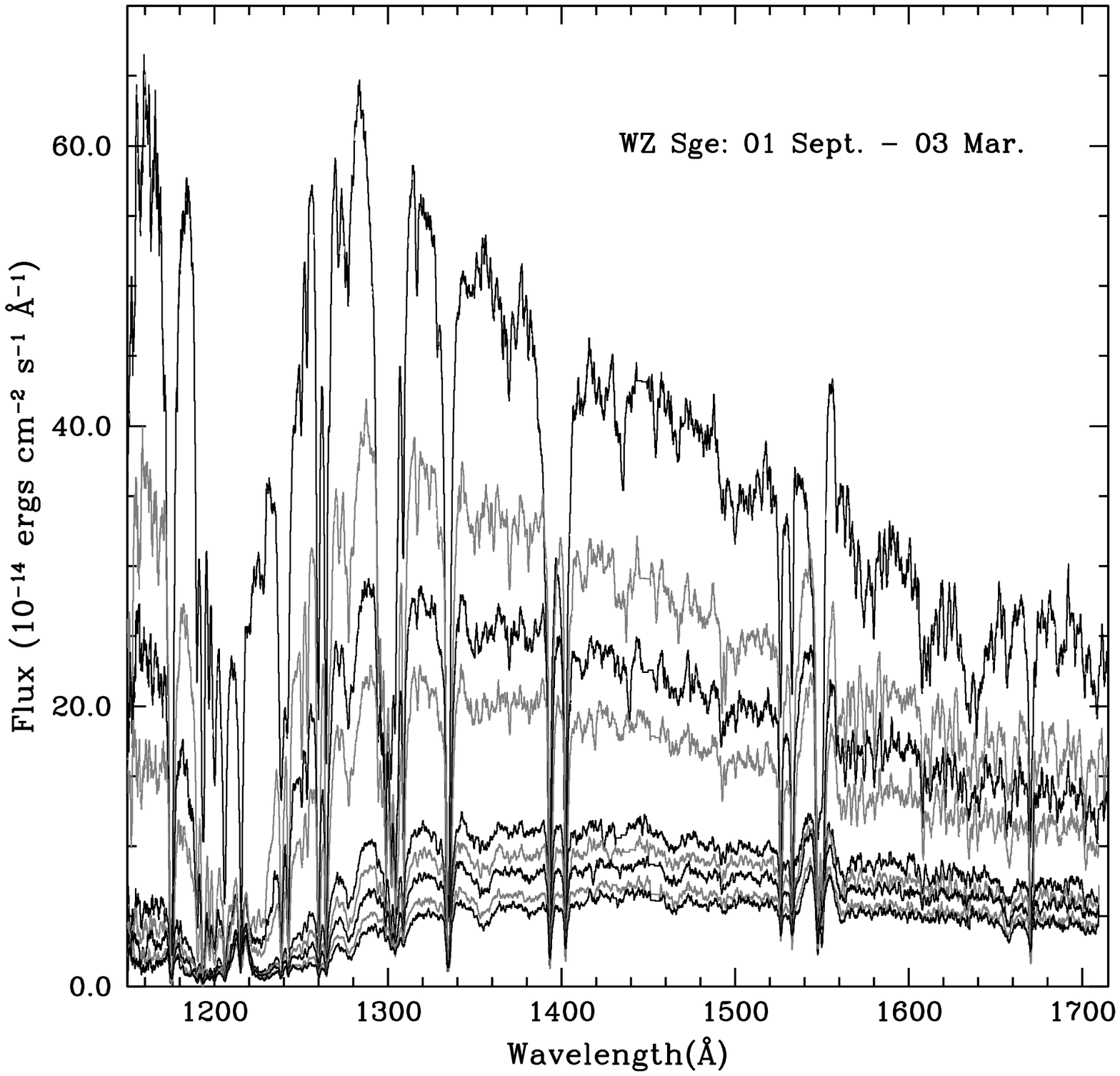}
\end{figure}

\begin{figure}
\plotone{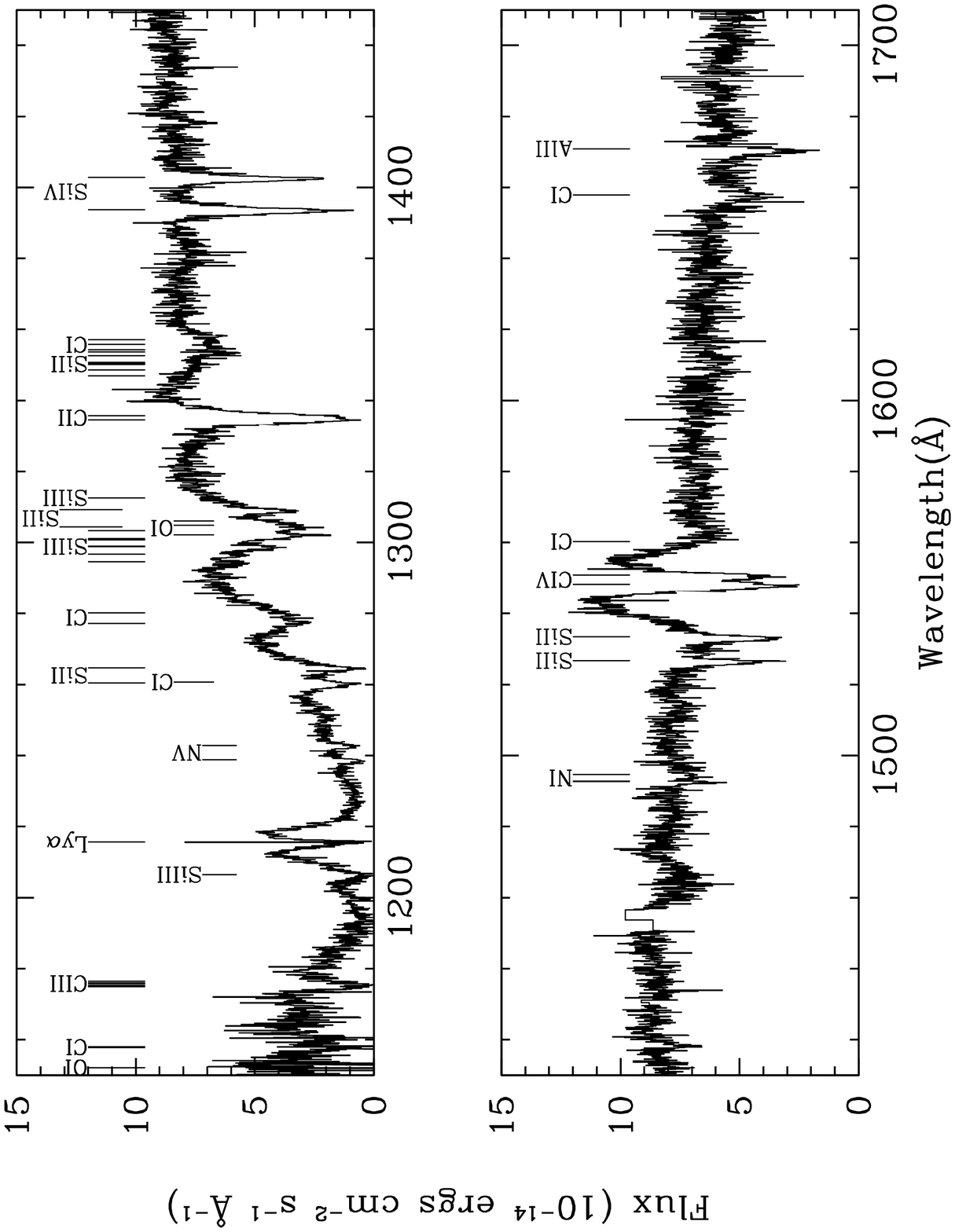}
\end{figure}

\begin{figure}
\plotone{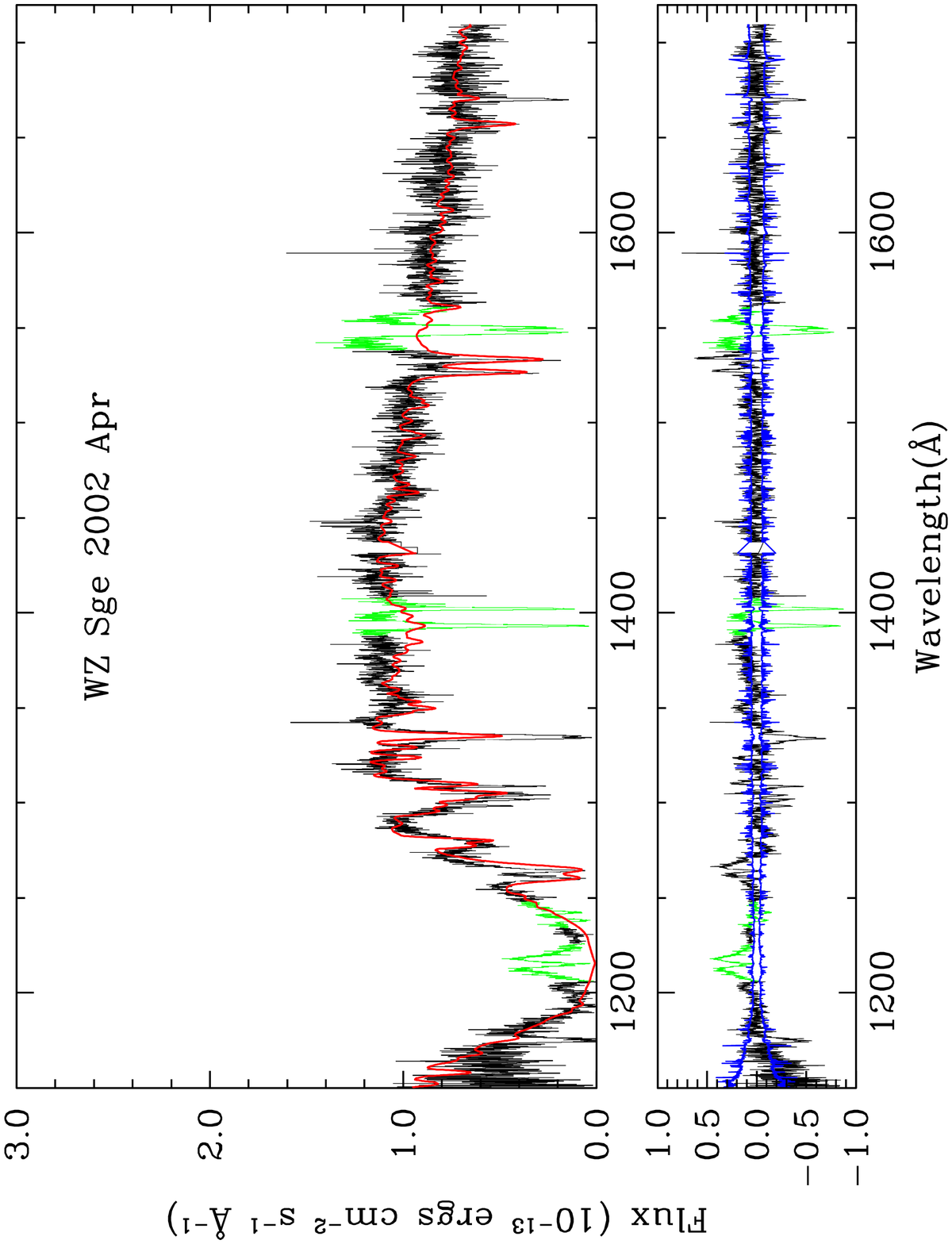}
\end{figure}

\begin{figure}
\plotone{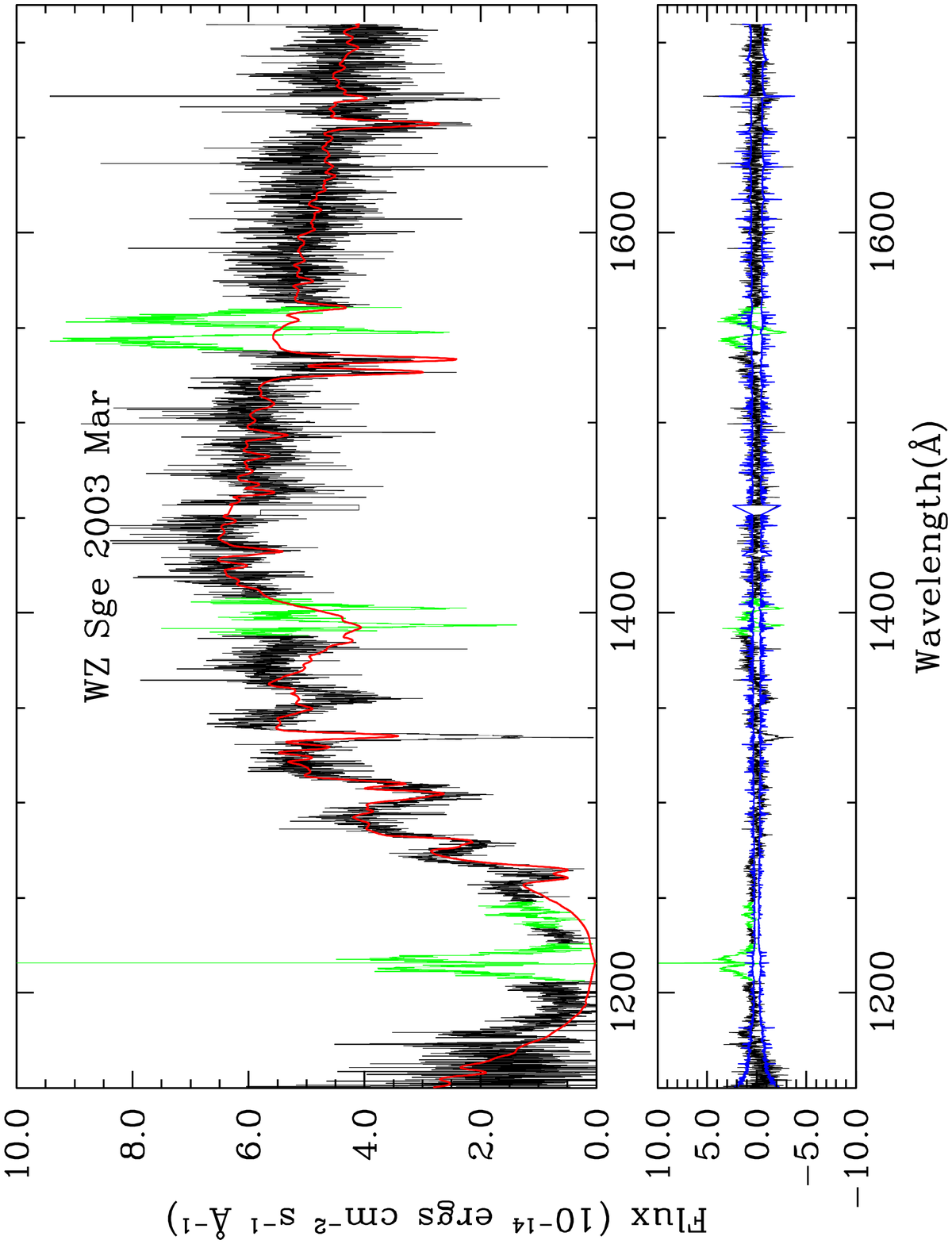}
\end{figure}

\begin{figure}
\plotone{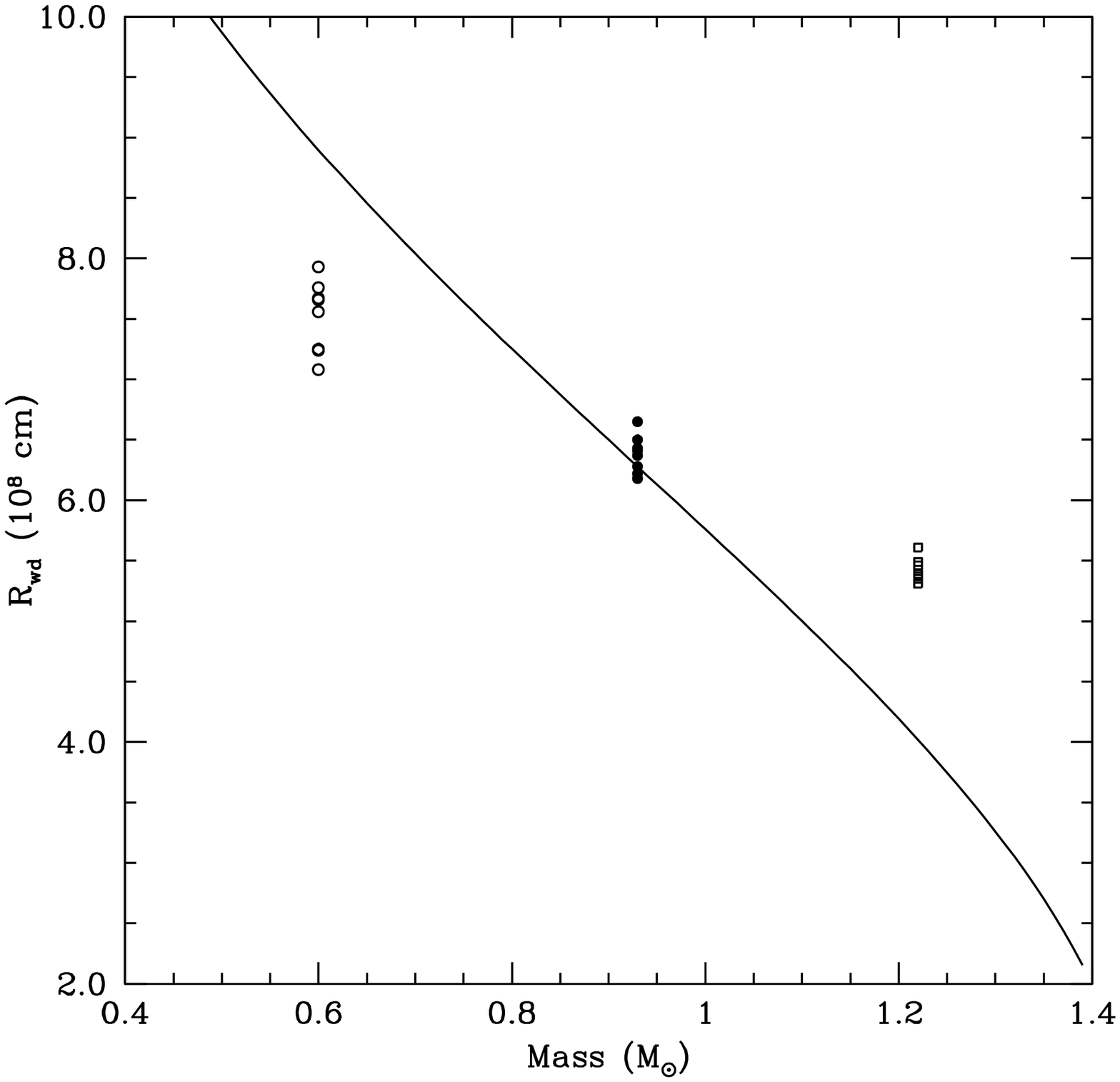}
\end{figure}

\begin{figure}
\plotone{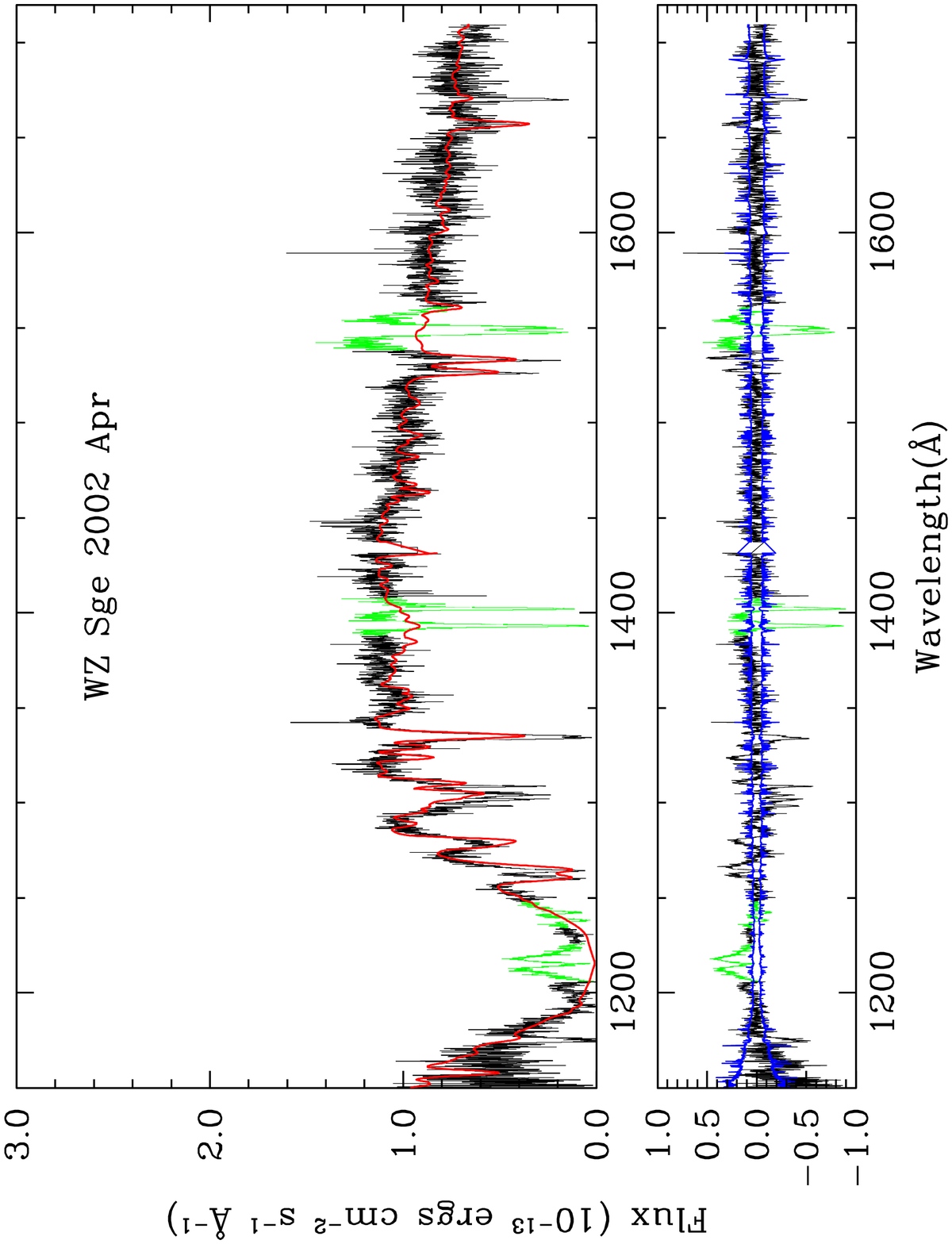}
\end{figure}

\begin{figure}
\plotone{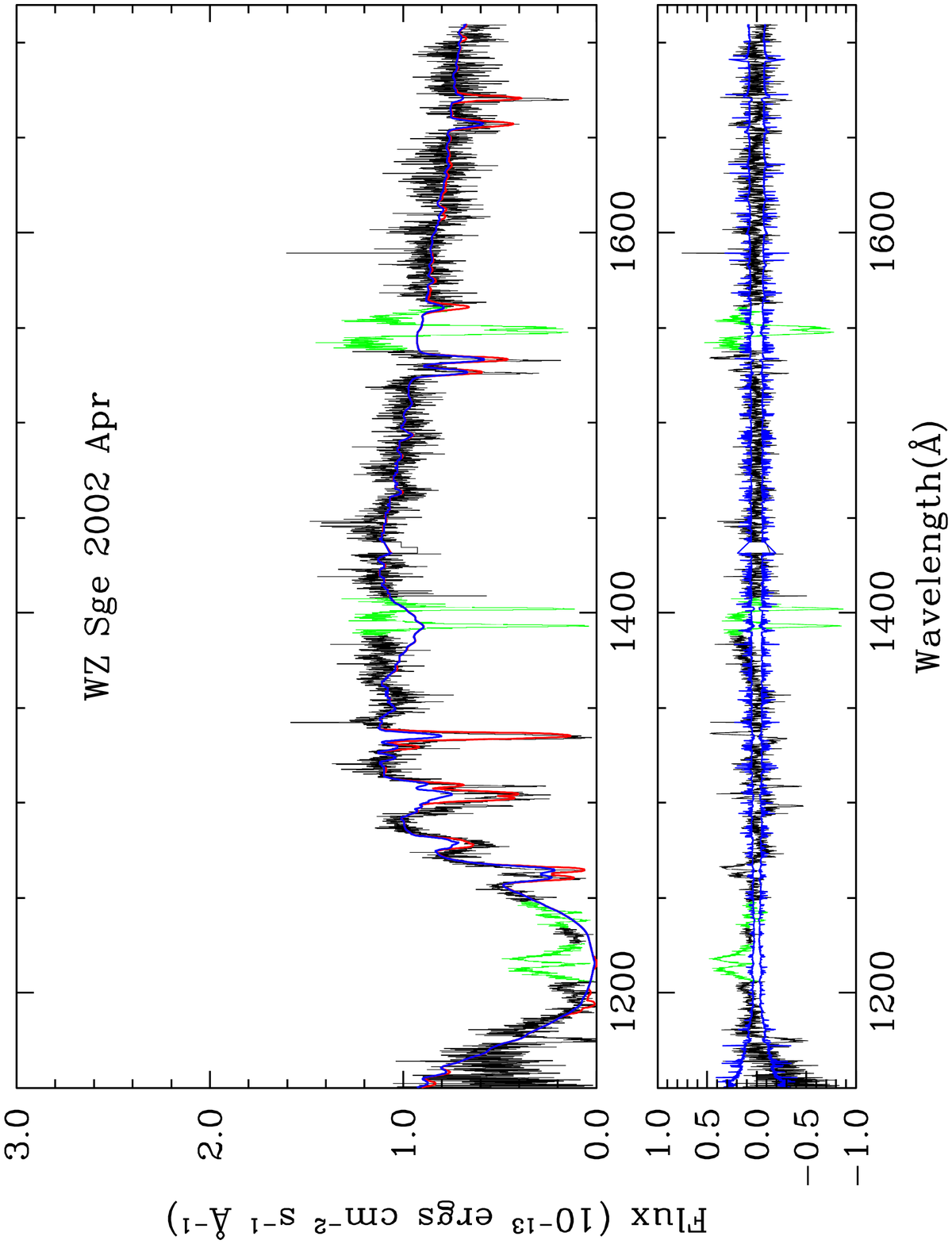}
\end{figure}

\begin{figure}
\plotone{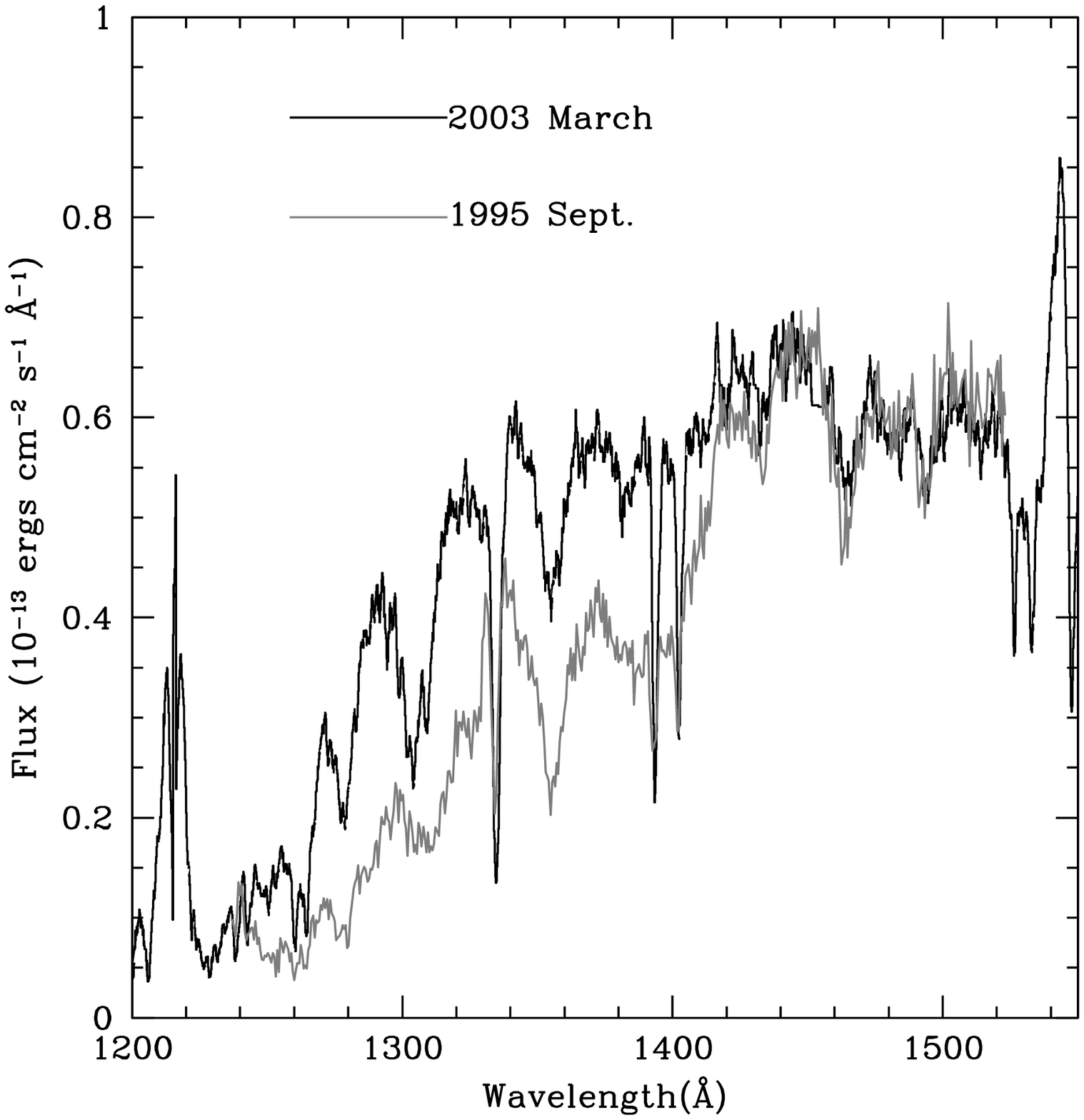}
\end{figure}

\begin{figure}
\plotone{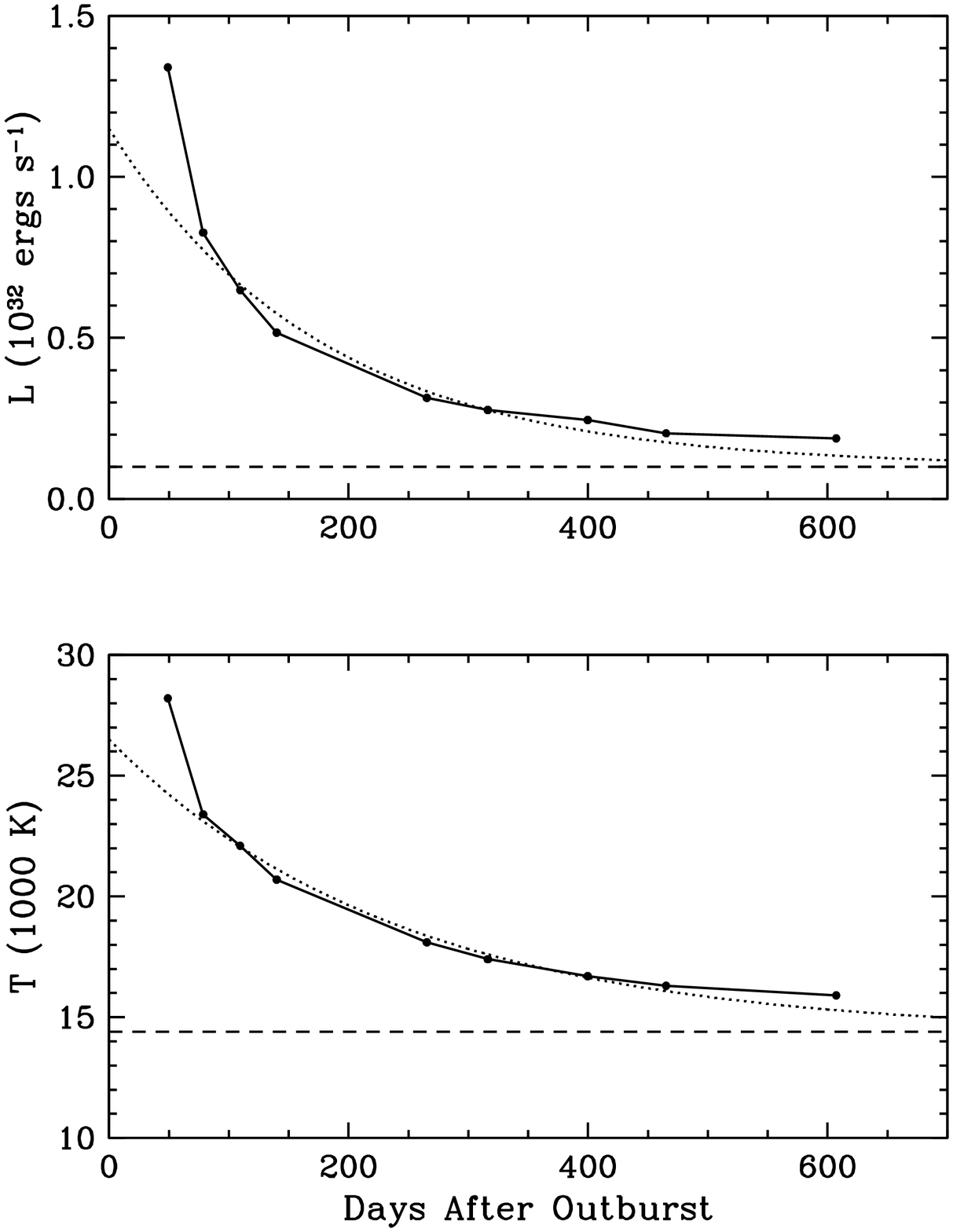}
\end{figure}

\begin{thebibliography}{}


\bibitem[Anderson(1988)]{anderson1988} Anderson, N.\ 1988, \apj,
325, 266


\bibitem[Cheng et al.(1997)]{cheng1997} Cheng, F.~H., Sion, E.~M., Szkody, P., \& Huang, M.\ 1997, \apjl, 484, L149








\bibitem[G{\" a}nsicke \& Beuermann(1996)]{gaensicke1996} G{\" a}nsicke,
B.~T.~\& Beuermann, K.\ 1996, \aap, 309, L47




\bibitem[Godon et al.(2003)]{godon2003} Godon, P., et al. 2003, ApJ,
submitted.

\bibitem[Hamada \& Salpeter(1961)]{hamada1961} Hamada, T.~\&
Salpeter, E.~E.\ 1961, \apj, 134, 683

\bibitem[Hameury, Lasota, \& Warner(2000)]{hameury2000} Hameury,
J., Lasota, J., \& Warner, B.\ 2000, \aap, 353, 244



\bibitem[Harrison et al.(2003)]{harrison2003}Harrison, T.~E., Johnson, J.~J., McArthur, B.~E., Benedict, G.~F.,
Szkody, P., Howell, S.~B., Gelino, D.~M. 2003, \aj, submitted



\bibitem[Horne et al.(1994)]{horne1994} Horne, K., Marsh, T.~R.,
Cheng, F.~H., Hubeny, I., \& Lanz, T.\ 1994, \apj, 426, 294

\bibitem[Hubeny(1988)]{hubeny1988}
Hubeny, I. 1988, Comput. Phys. Comm., 52, 103


\bibitem[Hubeny \& Lanz(1995)]{hubeny1995} Hubeny, I.~\& Lanz, T.\
1995, \apj, 439, 875

\bibitem[Kippenhahn \& Thomas(1978)]{kippenhahn1978} Kippenhahn, R.~\&
Thomas, H.-C.\ 1978, \aap, 63, 265

\bibitem[Knigge et al.(2002)]{knigge2002} Knigge, C., Hynes,
R.~I., Steeghs, D., Long, K.~S., Araujo-Betancor, S., \& Marsh,
T.~R.\ 2002, \apjl, 580, L151






\bibitem[Long et al.(1993)]{long1993} Long, K.~S., Blair, W.~P.,
Bowers, C.~W., Davidsen, A.~F., Kriss, G.~A., Sion, E.~M., \&
Hubeny, I.\ 1993, \apj, 405, 327


\bibitem[Long et al.(2003)]{long2003} Long, K.~S., Froning,
C.~S., G{\" a}nsicke, B., Knigge, C., Sion, E.~M., \& Szkody, P.\
2003, \apj, 591, 1172









\bibitem[Osaki, Meyer, \& Meyer-Hofmeister(2001)]{osaki2001}
Osaki, Y., Meyer, F., \& Meyer-Hofmeister, E.\ 2001, \aap, 370,
488



\bibitem[Patterson et al.(2002)]{patterson2002} Patterson, J.~et al.\
2002, \pasp, 114, 721

\bibitem[Pringle(1988)]{pringle1988} Pringle, J.~E.\ 1988, \mnras,
230, 587





\bibitem[Sion(1995)]{sion1995a} Sion, E.~M.\ 1995, \apj, 438, 876

\bibitem[Sion et al.(1995)]{sion1995} Sion, E.~M., Cheng, F.~H., Long, K.~S., Szkody, P., Gilliland, R.~L., Huang, M., \& Hubeny, I.\ 1995, \apj, 439, 957

\bibitem[Sion et al.(2003)]{sion2003} Sion, E.~M.~et al.\ 2003,
\apj, 592, 1137



\bibitem[Skidmore et al.(2000)]{skidmore2000} Skidmore, W., Mason, E., Howell, S.~B., Ciardi, D.~R., Littlefair, S., \& Dhillon, V.~S.\ 2000, \mnras, 318, 429



\bibitem[Slevinsky et al.(1999)]{slevinsky1999} Slevinsky, R.~J., Stys, D., West, S., Sion, E.~M., \& Cheng, F.~H.\ 1999, \pasp, 111, 1292


\bibitem[Spruit \& Rutten(1998)]{spruit1998} Spruit, H.~C.~\&
Rutten, R.~G.~M.\ 1998, \mnras, 299, 768

\bibitem[Steeghs et al.(2001)]{steeghs2001} Steeghs, D., Marsh, T.,
Knigge, C., Maxted, P.~F.~L., Kuulkers, E., \& Skidmore, W.\ 2001,
\apjl, 562, L145

\bibitem[Szkody et al.(2003)]{szkody2003}Szkody, P., G\"ansicke, B.~T., Sion, E.~M., Howell, S.~B., Cheng,
F.-H. 2003, \aj, 126, in press


\bibitem[Thorstensen (2003)]{thorstensen2003} Thorstensen, J. R.
2003, \aj, submitted

\bibitem[Welsh et al.(1997)]{welsh1997} Welsh, W.~F., Skidmore, W., Wood, J.~H., Cheng, F.~H., \& Sion, E.~M.\ 1997, \mnras, 291, L57

\bibitem[Welsh et al.(2003)]{welsh2003} Welsh, W.~F., Sion,
E.~M., Godon, P., Gansicke, B.~T., Knigge, C., Long, K.~S., \&
Szkody, P.\ 2003, \apj, in press

\end{thebibliography}
\end{document}